\def\lesssim{{_ <\atop{^\sim}}}
\def\grtsim{{_ >\atop{^\sim}}}
\def\ap3m{AP$^3$M}
\def\LCDM{$\Lambda$CDM}
\def\LWDM{$\Lambda$WDM}
\def\hkpc{$h^{-1}{\ }{\rm kpc}$}
\def\hMpc{$h^{-1}{\ }{\rm Mpc}$}
\def\hMsun{$h^{-1}{\ }{\rm M_{\odot}}$}

\def\nbody{$N$-body}
\def\c15{$c_{\rm 1/5}$}

\newcommand{\Eq}[1]{Eq.~(\ref{#1})}
\newcommand{\Sec}[1]{Section~\ref{#1}}
\newcommand{\Fig}[1]{Fig.~\ref{#1}}
\newcommand{\mlapm}{\texttt{MLAPM}}

\def\ea{et~al.~}                            

\def\lesssim{\mathrel{\hbox{\rlap{\hbox{\lower4pt\hbox{$\sim$}}}\hbox{$<$}}}}
\def\gtrsim{\mathrel{\hbox{\rlap{\hbox{\lower4pt\hbox{$\sim$}}}\hbox{$>$}}}}

\newcommand{\AAA}[3]    {\mbox{#3,~A\&A,~\textbf{#1},~#2}}

\newcommand{\ApJ}[3]    {\mbox{#3,~ApJ,~\textbf{#1},~#2}}

\newcommand{\ApJL}[3]   {\mbox{#3,~ApJ~Lett.,~\textbf{#1},~#2}}

\newcommand{\MNRAS}[3]  {\mbox{#3,~MNRAS,~\textbf{#1},~#2}}

\newcommand{\PhRevL}[3] {\mbox{#3,~Phys.~Rev.~Lett.,~\textbf{#1},~#2}}

\newcommand{\astroph}[1]{\mbox{\texttt{astro-ph/#1}}}

\documentstyle[epsfig]{mn}

\begin{document}

\title{Top-Down Fragmentation of a Warm Dark Matter Filament}

\author[Knebe A., Devriendt J.E.G., Gibson B.K., Silk J.]
       {Alexander Knebe$^1$, Julien E. G. Devriendt$^2$, Brad K. Gibson$^1$,
        Joseph Silk$^2$\\        
       {$^1$Centre for Astrophysics \& Supercomputing,
        Swinburne University, P.O. Box 218, Mail \# 31,
        Hawthorn, Victoria, 3122, Australia}\\
       {$^2$Astrophysics, Oxford University, Keble Road, Oxford, OX1 3RH, UK}}

\date{Received ...; accepted ...}

\maketitle

\begin{abstract}
We present the first high-resolution \nbody~simulations of the
fragmentation of dark matter filaments. Such fragmentation occurs in
top-down scenarios of structure formation, when the dark matter is
warm instead of cold.  In a previous paper (Knebe~\ea 2002, hereafter
Paper~I), we showed that WDM differs from the standard Cold Dark
Matter (CDM) mainly in the formation history and large-scale
distribution of low-mass haloes, which form later and tend to be more
clustered in WDM than in CDM universes, tracing more closely the
filamentary structures of the cosmic web.  Therefore, we focus our
computational effort in this paper on \textit{one} particular filament
extracted from a WDM cosmological simulation and compare in detail its
evolution to that of the same CDM filament.  We find that the mass
distribution of the halos forming via fragmentation within the
filament is broadly peaked around a Jeans mass of a few $10^9$
M$_\odot$, corresponding to a gravitational instability of smooth
regions with an overdensity contrast around 10 at these redshifts.
Our results confirm that WDM filaments fragment and form
gravitationally bound haloes in a top-down fashion, whereas CDM
filaments are built bottom-up, thus demonstrating the impact of the
nature of the dark matter on dwarf galaxy properties.

\end{abstract}


\section{Introduction}

The Cold Dark Matter crisis on small scales is far from being
resolved.  First, the highest resolution simulations available still
favour "cuspy" galaxy haloes (Power~\ea 2003) with an inner slope for
the density profile $\approx -1.2$, whereas high resolution
observations of low surface brightness galaxies are best fit by halo
"cores" (de Block \& Bosma 2002; Swaters~\ea 2003).  Second, the
dearth of low luminosity galaxy satellites found \nbody\ simulations
remains, although squelching of dwarfs due to the reionisation of the
universe at high redshift may provide a solution (e.g. Benson~\ea
2002; Somerville 2002). However, it is unclear that such a process is
the definite solution, because the low-mass dark satellite haloes left
over still interact gravitationally with the luminous galaxies,
exchanging energy and angular momentum with them. Such interactions
could have a significant impact on these galaxies' properties, heating
their disks and reducing their sizes.  Direct observational probing of
dark matter halo substructure through anomalies in the flux ratio of
multiple-image quasar lenses could in principle determine whether the
scenario previously suggested is the correct one (Metcalf \& Madau
2001; Dalal \& Kochanek 2002), but there still are major difficulties
to overcome in order to obtain reliable estimates of the amount of
mass locked in dark substructures (Chen~\ea 2003).  

An alternate mechanism to solve the satellite problem, which does not
suffer from these side effects, is to reduce the power of dark matter
fluctuations on small scales.  This can be accomplished by changing
the nature of the dark matter, (e.g. Kaplinghat, Knox, \& Turner 2000;
Spergel \& Steinhardt 2000; Colin~\ea 2000; Bode, Ostriker~\& Turok
2001; Avila-Reese~\ea 2001), by advocating an inflationary model with
broken-scale invariance (Kamionkowski \& Liddle 2000) or by simply
assuming an ``unusual'' dip in the primordial CDM power spectrum on
small scales (Little, Knebe~\& Islam 2003).  The latest constraints
from polarisation measurements of the Cosmic Microwave Background
Radiation by WMAP (Bennett et al., 2003; Spergel et al., 2003; Kogut
et al., 2003) seem to exclude such models, with the major caveat that
the current theoretical understanding of the objects that produce the
first ionising photons is at best crude. As an example, a reionisation
redshift of 11 (on the lowest side of the WMAP published 95 \%
confidence level) is still reached in a WDM model for which the warmon
mass is about 1 keV, provided the objects responsible for
photoionising the universe at this redshift emit on average between
100 and 1000 times more ionising photons per baryon than normal stars,
as is possible with a top--heavy IMF dominated by accreting black
holes (Schneider~\ea 2002). Moreover, the constraints become less
restrictive as the mass of the warmon increases. 

More sophisticated numerical simulations of high-redshift WDM--like
universes, combined with a better understanding of the reionising
photon emitters are required in order to set definitive constraints on
the nature of the dark matter from polarization measurements. This
important point set aside, previous simulations (see Paper~I; Bode,
Ostriker~\& Turok 2001; Avila-Reese~\ea 2001) have indicated that
deviation of WDM models from the CDM hierarchical structure formation
scenario should be more pronounced in low density regions
(e.g. filaments, voids). More specifically, in these simulations, most
of the halos close to the resolution limit (i.e. less than 100
particles) seemed to be forming at smaller redshifts than in the
standard CDM models, and in a top-down fashion through the break-up of
filamentary structures.  This phenomenon is of prime interest for
galaxy formation and evolution models as it predicts that dwarfs
properties strongly depend on the nature of the dark matter.  It also
gives rise to a series of interesting questions: how much do filaments
fragment?  When do they fragment the most? What is the mass
distribution of the fragments?  To tackle this issue we identified one
of the filaments in the previous CDM and WDM low resolution runs
presented in Paper~I and resimulated it with a mass resolution 64
times higher than the original.

This paper proceeds as follows:
in \Sec{Model} we briefly give the model parameters and describe the \nbody-simulations which we use 
in \Sec{Analysis} to calculate the properties of gravitationally
bound halos found in the filament and of the filament itself. 
Finally, we present our conclusions  in \Sec{Conclusions}.

\section{$N$-BODY simulations} \label{Model}

The \LWDM\ and the \LCDM\ cosmological parameters used in this paper are the same 
as those of the \LWDM2 and \LCDM\ runs presented in Paper~I,  
namely, $\Omega_0 = 1/3$, $\lambda_0 = 2/3$, $\sigma_8=0.88$,
$h=2/3$, and $m_{\rm WDM} = 0.5$keV.
We note that they are in good agreement with the latest results from 
WMAP (Bennett~\ea 2003).

   \begin{figure}
    \centerline{\psfig{file=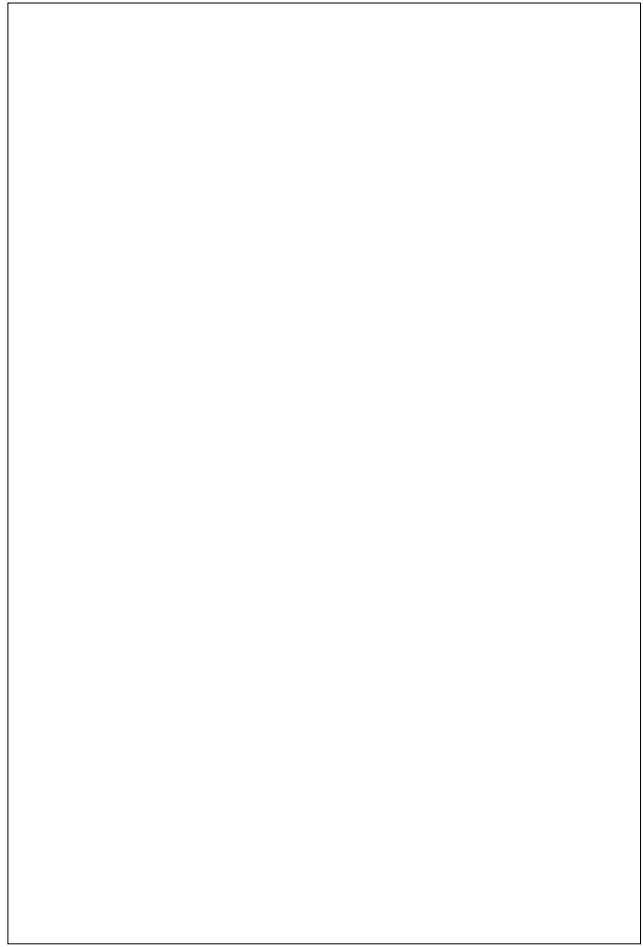,width=\hsize}}
    \caption{Gray-scaled image of the logarithm of the density field 
    around the filament at redshifts $z=1$ (upper panel), $z=3$ (middle panel) 
    and $z=5$ (lower panel).}  
    \label{WDMfila}
   \end{figure}

As outlined in Paper~I, initial conditions for our lower resolution
simulations were generated by packing particles from higher resolution
initial conditions into groups. For the simulations presented in this
paper, we simply identified a filament at $z=0$ in our 25~\hMpc~ low
resolution box, tracked back in time the Lagrangian volume its
particles occupied at $z=35$, and unpacked the low-resolution
particles into high-resolution particles in that region.  In practice,
our resimulation of the filament resulted in a 64-fold increase in
mass resolution from $m_p \simeq 7
\times10^8$\hMsun\ (128$^3$ particles) in the low resolution runs to $m_p \simeq
10^7$\hMsun\ (512$^3$ particles) in the runs presented here.  Each of
our \LWDM\ and \LCDM\ filaments contains over 3 million high
resolution particles and our force resolution is of the order 3
\hkpc\ in the densest regions.  The total
length of a filament can be represented by the dimensions $x_b=7$,
$y_b=4$, $z_b=6$ (in \hMpc) of the rectangular box enclosing it.  We
took "snapshots" of the simulations at the following redshifts: $z$=5,
4.5, 4, 3.5, 3, 2.5, 2, 1.75, 1.5, 1.25, 1, 0.5, and 0.  These
re-simulations were undertaken using the publicly available Adaptive
Mesh Refinement code
\texttt{MLAPM} (Knebe, Green~\& Binney, 2001).

Particle groups were identified using the Bound Density Maxima (BDM,
Klypin~\& Holtzman 1997) method, in keeping with Paper~I.  The BDM
code identifies local overdensity peaks by smoothing the density field
on a particular scale of the order of the force resolution. We adopted
a smoothing of 10\hkpc\ for the runs under investigation. These peaks
are prospective halo centres. For each of these halo centres we step
out in (logarithmically spaced) radial bins until the density reaches
$\rho_{\rm halo}(r_{\rm vir}) =
\Delta_{\rm vir} \rho_b$, where $\rho_b$ is the background density
and $\Delta_{\rm vir}\approx 340$.  This defines the outer radius
$r_{\rm vir}$ of the halo. Once we know the outer radius we are able
to calculate internal properties of the halos. Note that during this
group identification procedure we also check if each individual
particle truly belongs to the halo by comparing its velocity to the
local value of the escape velocity. We exclude unbound particles from
the haloes. This iterative method ensures that we only retain
gravitationally bound objects.

\section{Results} \label{Analysis}

\subsection{Dark Matter Density Field} \label{DensityField}

We begin our investigation by looking at the large-scale structure density 
field within and around the filament. 
\Fig{WDMfila} shows the particle distribution for
both models with each particle being grey-scaled proportionally to the
logarithm of the local (over-)density. A striking feature of 
\Fig{WDMfila} is the marked granularity of the density field in 
the CDM model at all redshifts in contrast with WDM. This effect
reflects the lack of filtering of power on small scales in CDM as
opposed to WDM.  However, while the WDM filament is obviously
different from its CDM counterpart at redshift $z$=5, they are
very similar at $z$=1. This similarity is all the more remarkable
that the logarithmic scale used to plot the
density fields shown in \Fig{WDMfila} artificially enhances the
density contrast, thus highlighting differences between the two 
universes that would have gone unnoticed had the scale been linear.

A further difference of note is the \textit{appearance} of individual
(low-mass) halos during the course of the WDM filament evolution,
whereas dense objects only seem to be merging in the CDM filament,
therefore \textit{decreasing} steadily in number.  

We have not presented results at redshift $z=0$ because the picture at
this epoch is qualitatively similar to that at reshift $z=1$: most of
the interesting processes leading to different looking WDM and CDM
filaments occur at high redshifts ($z>1$).

\subsection{Halo Abundance in the Filament} \label{mass}

   \begin{figure}
    \centerline{\psfig{file=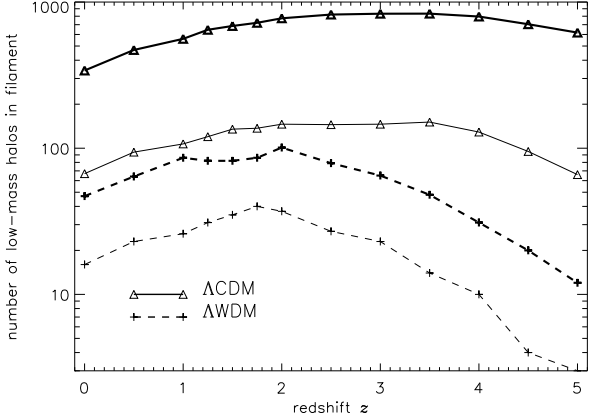,width=\hsize}}
    \caption{Number of gravitationally bound halos located with
             the filament as a function of redshift.
             We employed a mass cut of $M>10^{9}$\hMsun\ (thick lines)
             and $M>10^{10}$\hMsun\ (thin lines), respectively.}  
    \label{abundance}
   \end{figure}

   \begin{figure}
    \centerline{\psfig{file=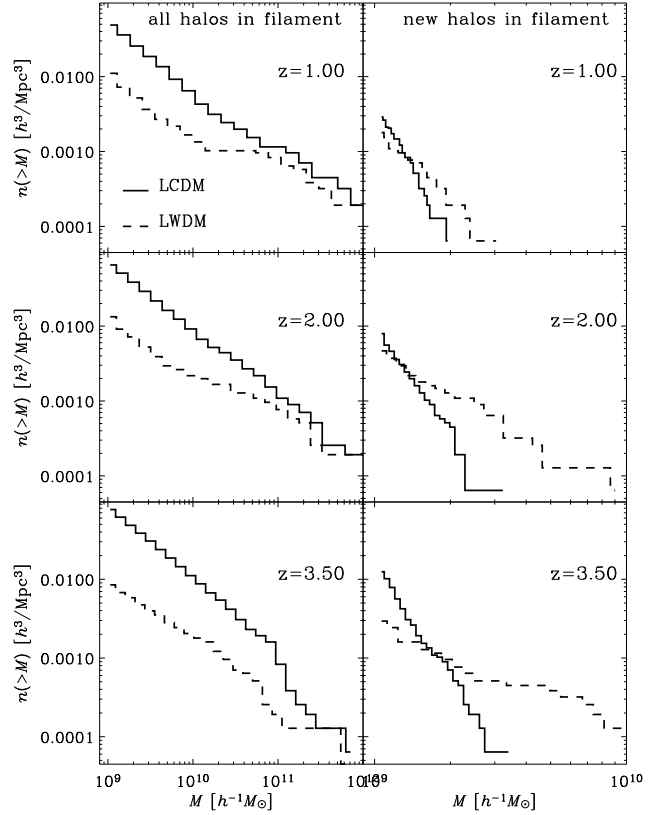,width=\hsize}}
    \caption{Cumulative mass functions of all halos located
             within the filament (left panel) along with
             the mass function of 'new' halos as defined
             in the text (right panel).}  
    \label{totmassfun}
   \end{figure}

To quantify the discussion in Section~\ref{DensityField}, we compute
how the number of halos within the filament evolves with redshift.
This number is plotted in
\Fig{abundance}, and contains all gravitationally bound particle
groups above $M > 10^{9}$\hMsun\ (more than 100 particles -- thick
lines) and $M > 10^{10}$\hMsun\ (more than 1000 particles -- thin
lines), respectively.  The former simply corresponds to the integral
of the mass function over the entire halo mass range available given
our resolution. 

\Fig{abundance} demonstrates that independent of redshift, there are more halos
in the CDM filament (as already expected from \Fig{WDMfila}) than in
the WDM filament.  However, the ratio $N_{\rm halo}^{\rm CDM} / N_{\rm
halo}^{\rm WDM}$, which appears to be nearly independent of the mass
cut employed, drops significantly from about 50 at $z=5$ to
approximately 7 at $z=2$. This ratio is a factor two smaller for the
higher mass-cut of $10^{10}$\hMsun.  The depletion of halos in both
models from $z=2$ to $z=0$ is readily explained when we consider that
very few objects are created within the filament at low redshifts (see
\Fig{newhalo}), and therefore, mergers must be the dominant process in
determining the number of halos in that redshift range.  Indeed, as
can be seen in \Fig{WDMfila} haloes stream along the web-like
structures to eventually become subsumed within larger objects (galaxy
groups or clusters) located at the intersections of the filamentary
network.

\Fig{totmassfun} displays the cumulative mass function of
all halos identified within the filament (left panel) along with the
mass function of 'new' halos (right panel). The definition of a 'new'
halo is given in Section~\ref{NewHalosInFila} where
\Fig{totmassfun} is analysed in more detail, but for the moment one can simply 
view them as haloes which did not have any progenitor at the previous
output. At this point of the analysis, we stress that the cumulative
mass function of {\em all} halos in the CDM filament closely follows a
power-law (with a slope close to -1) whereas in the WDM case one
observes a deficit of low-mass objects, in agreement with the results
presented in Paper~I.

\subsection{Mass of the Filament} \label{MassOfFila}

   \begin{figure}
    \centerline{\psfig{file=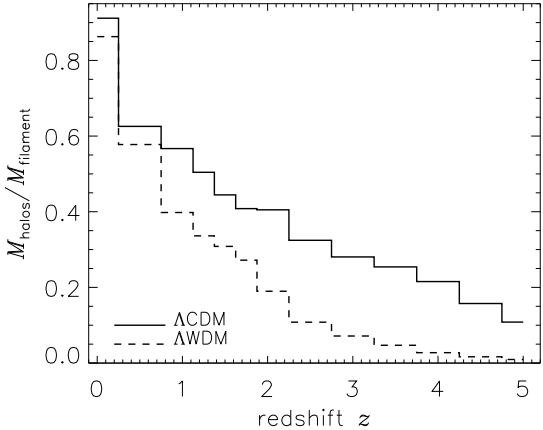,width=\hsize}}
    \caption{Fraction of mass within gravitationally bound halos
             of mass $M>10^9$\hMsun\ as a function of redshift.}  
    \label{mfrac}
   \end{figure}

We next addres the question of the fraction of the filament material
which gets locked up in galactic halos and how much remains smoothly
distributed in the filament. To answer this question, we calculate the
ratio of particles belonging to halos with $M > 10^{9}$\hMsun\
(identified using the BDM code, cf. \Fig{abundance}) to the total
number of particles in the filament.  This requires a definition for
the boundary of the filament which follows from a previous definition
for voids: Einasto~\ea (1998) define \textit{void particles} as having
$\rho_p < \rho_b$ where $\rho_p$ is the local density at the particle
position and $\rho_b$ the cosmological background density. We adopt
the same prescription and call those particles that follow $\rho_p >
\rho_b~$\textit{filament particles}. The density at the particle position 
is interpolated from the density field calculated on a regular 256$^3$
grid; as we restrict ourselves to only high-resolution particles we
are at all times confined to the filament region (by definition).

The results of this analysis are plotted in
\Fig{mfrac}. This figure highlights the overabundance of filament halos
in CDM with respect to WDM, as the fraction of material in particle
groups is at all times higher in the former than in the latter. Only
at $z \lesssim 1$ does the ratio $M_{\rm halos}/M_{\rm filament}$
become comparable in both models, which confirms the similarity of
these structures previously noticed at $z=1$ (cf. \Fig{WDMfila}).  We
note however that from \Fig{WDMfila} alone, one might have claimed
that most of the material should already be in haloes at $z=5$ in the
CDM filament, but \Fig{mfrac} argues against it, as only about 10 \%
of the particles belong to haloes more massive than $M >
10^{9}$\hMsun, for this model, at this redshift. Of course, in the CDM
case, this percentage is quite sensitive to the mass cut-off adopted, in
the sense that if we had chosen to consider haloes composed of 50
particles or more, it would have increased to about 20 \%.
Nevertheless, this implies that the gravitational field within the
filament is still dominated by the smoothly distributed dark matter
component up to redshifts $z \simeq$ 1, both for the CDM and WDM
filaments, and that accretion and merging are indeed important since
the total number of halos decreases below $z=2$ in both structures
whereas the mass of each filament which is locked up within those
haloes increases.

\subsection{New Halos in the Filament} \label{NewHalosInFila}

\begin{figure}
    \centerline{\psfig{file=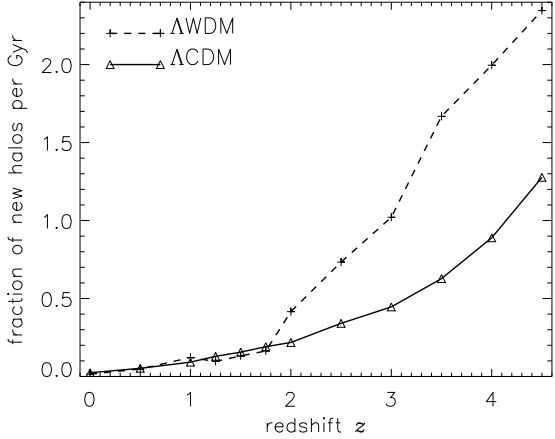,width=\hsize}} 
    \caption{Fraction of halos per Gyr appearing in the filaments
             and for which no progenitor could be found at the
             previous redshift (new haloes). Note that the fraction 
             becomes greater than unity at high redshift. This simply 
             means that had the actual halo formation rate been 
	     the same at earlier times, more halos would have been 
             formed than are currently present in the filament.}
    \label{newhalo} 
   \end{figure}

This leads us to the fundamental question of this study, namely {\it
how} and {\it when} does the WDM filament fragment to form haloes?  In
other words, how important is this "top-down" fashion of forming haloes
compared to the more traditional hierarchical "bottom-up" scenarios of
CDM-like models?  To tackle this issue, we constructed the merger tree
of each individual halo between consecutive redshift outputs and
tagged a halo as 'new' at a given redshift when no progenitor could be
found at the previous output. The requirement for a progenitor to be
identified as such was that it had to contain a minimum of 100
particles or 1/10 of the actual halo particles depending on which of
these two numbers was the smallest. If such a progenitor was found the
halo was marked 'old'.
\Fig{newhalo} shows the fraction of 'new' halos formed per unit time as a function of
redshift. As can be seen from this figure, the interesting redshift
range for this analysis is from $z=4.5$ to $z=2$, as one observes a
significant halo formation rate within the WDM filament during this
time period (cf. \Fig{newhalo}).

In order to obtain results not affected by spurious formation or
evaporation of haloes between timesteps, we have set a lower mass
limit of $10^{9}$\hMsun\ (which translates into 100 particles). This
means that if a halo is less massive than this cut-off it has been
neglected in our analysis.  However, this does not imply that its
progenitor(s) had to contain 100 particles. The mass limit for the
progenitor(s) is the minimum of 100 and 1/10 of the actual halo
particles and hence can be as low as 10 particles (minimal mass for
which a bound object can reasonably be detected). We emphasize that in
the extreme case where a 10 particle progenitor of a halo is
identified, it only affects the 'tag' of the halo currently under
consideration, meaning that it is classified as 'old' instead of
'new'. More specifically, if such a progenitor turns out to be
unphysical (a mere collection of 10 particles which happen to pass by
one another quite slowly) it biases the results towards
\textit{fewer} new halos being created in the filament. In this sense, 
the curves plotted in \Fig{newhalo} are lower limits on the true
number of new haloes in the mass range $10^{9}$\hMsun\ and above, as
we have made sure that the halos marked as 'new' were not a part of
\textit{any} structure at the previous redshift. \textit{\Fig{newhalo} alone
strongly suggests that the bulk of the halo population is forming
between $z=4.5$ and $z=2.5$ within the WDM filament whereas the same
figure points at the majority of halos being already in place in the
CDM structure by $z=4$}. To be more explicit, this figure shows that
the WDM halo formation rate at $z=3$ is still high enough that it
could account for all halos present at this redshift, had it been
constant in the past.  But we also know from \Fig{abundance} that the
maximum number of WDM halos is broadly peaked around $z=2$, and from
\Fig{newhalo} that the halo formation rate is not constant but steeply
increasing with redshift for $z>2$.  Furthermore, \Fig{newhalo} and
\Fig{merger} tell us that the halo merger rate is negligible compared
to the halo formation rate at $z \geq 2$. Therefore, we are forced to
conclude that most halos were indeed 'born' in the WDM filament
between $z=5$ and $z=2$. The same analysis, applied to the CDM
filament leads to the claim that most halos are in place before $z=5$.

\begin{table}
\caption{Characteristics of the 749 particles that end up forming
 the gravitationally bound halo at $z=2$ shown in \Fig{WDMfragment}.}
\label{fragment}
\begin{tabular}{lccc}

redshift   & $\langle \rho \rangle/\rho_b$ & $\langle r \rangle$/(\hkpc) & $\Delta_{\rm vir}$\\ 
\hline \hline
$z=2.0$    &         278                   &        35                   &   210\\
$z=2.5$    &         200                   &        96                   &   208\\
$z=3.0$    &         190                   &       137                   &   207\\
$z=3.5$    &         82                    &       165                   &   207\\

\end{tabular}
\end{table}

   \begin{figure}
    \centerline{\psfig{file=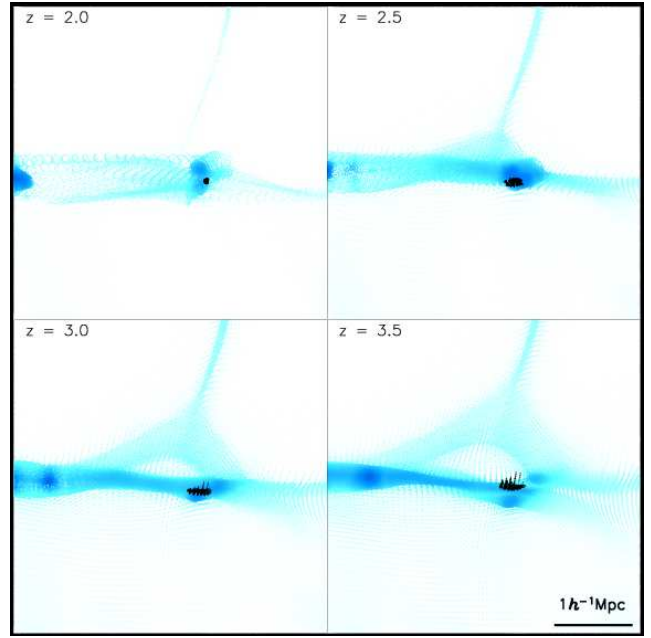,width=\hsize}} 
    \caption{Example of a halo fragmentation within the WDM filament. 
             All 749 particles which end up in the virial radius of the bound 
             halo at $z=2$ are shown in black. The grey-shaded area 
             represents the density field within a filament slice. The thickness 
             of this slice is equal to the maximal distance between halo 
             particles at any time.}
    \label{WDMfragment} 
   \end{figure}

This answers the question as to {\it when} the halos form out of the
WDM filament, but
\Fig{WDMfragment} shows an example of {\it how} such a fragmentation occurs.
The sequence of panels in that figure traces back in time all the
particles found in a typical new halo of mass $M \approx 7.5 \times
10^{9}$\hMsun\ at redshift $z=2$ (i.e. a halo which was identified to
be a 'new' gravitationally bound object at this redshift using the BDM
algorithm).  It is obvious from \Fig{WDMfragment} that particles which
end up within the virial radius of this halo at $z=2$ are part of the
filamentary structure until the very moment when the collapse takes
place. The quantitative information obtained by tracing the particles 
pertaining to the final bound halo back in time and 
provided in Table~\ref{fragment} backs up this qualitative analysis. 
In particular, we computed in Table~\ref{fragment} the averaged local 
density (in terms of the background density) at all
749 particle positions, as well as the mean extent of the region
occupied by those particles. As a reference we also listed the virial
overdensity value $\Delta_{\rm vir}$ for each of the redshifts under
consideration in the last column of Table~\ref{fragment}.

To obtain a physical explanation for the fragmentation, we can calculate
how the Jeans mass, $M_J$, varies with redshift in the WDM filament
prior to its fragmentation, assuming its density field to be close to
homogeneous on scales larger than the Jeans scale $\lambda_J$.  These
masses and scales are given by the following formulae (e.g. Binney~\&
Tremaine 1987):

\begin{equation} \label{Mjeans}
 M_J = \frac{\pi^{5/2}}{6 G^{3/2}} \frac{\sigma_v^3}{\sqrt{(1+\delta) \rho_b}}
\end{equation}

\begin{equation} \label{Ljeans}
 \lambda_J = \sqrt{\frac{\pi}{G \rho_b (1+\delta)}} \sigma_v
\end{equation}

\noindent
where $\rho \equiv (1+\delta) \rho_b$ is the density of the filament
and $\sigma_v$ its averaged velocity dispersion.  In the previous
equations, $\delta$ stands for the (over-)density contrast within the
filament with respect to the mean density of the background universe,
$\rho_b$, at a given redshift.  For our cosmological model,
\Eq{Mjeans} and \Eq{Ljeans} become

\begin{equation} 
 M_J = 5.1 \times 10^7 (1+z)^{-3/2} (1+\delta)^{-1/2}
\left(\frac{\sigma_v}{\mathrm{km/s}}\right)^3 \mathrm{M_\odot}
\end{equation}

\begin{equation} 
 \lambda_J = 0.13 \times (1+z)^{-3/2} (1+\delta)^{-1/2} \frac{\sigma_v}{\mathrm{km/s}} \mathrm{Mpc}
\end{equation}

A rough estimate of the mass of the fragments can be calculated by
assuming that the average density contrast of the filament, $\delta
\approx 10$ is a slowly varying function of redshift between $z=5$ and
$z=2$, and that typical velocity dispersions are of the order of 20
km/s\footnote{This value is based on the averaged velocity dispersion
of particles measured in the simulated filaments at local
overdensities of around 10.}.  We then get $ 8 \times 10^{9}
\mathrm{M_\odot} \leq M_J \leq 2
\times 10^{10} \mathrm{M_\odot}$, to be compared to the mass functions
for new haloes given in \Fig{totmassfun} (right panel) at redshifts
3.5, 2 and 1.  We observe that the masses of the new WDM halo are
actually close to these estimated values for the Jeans mass, thus
supporting the assumption that they do indeed form through a
fragmentation of the filament caused by gravitational instability.
Now, the same range of values for $\delta$, $z$ and $\sigma_v$ yields
$ 53 \; \mathrm{kpc} \leq \lambda_J \leq 150 \; \mathrm{kpc}$ which
validates our assumption for the homogeneity of the WDM filament
density field on these scales since the free--streaming length $R_f$
of a 0.5 keV particle is $R_f \simeq 300 \; \mathrm{kpc}$
corresponding to a free streaming mass $M_f \simeq 4 \times 10^{10}
\mathrm{M_\odot}$.  Note that this is not the case for the CDM
filament as the free-streaming scale associated with the cold particle
is much smaller than the Jeans length, and therefore we cannot
consider the density field as homogeneous on these scales.  Moreover,
one consideration regarding the simulation technique needs to be borne
in mind: our \nbody\ code \mlapm\ achieves high spatial resolution
primarily in high density regions.  This means that the force
resolution can smooth out the density field on scales above the Jeans
mass in lower density regions, with the result that one would then
tend to identify haloes as 'new' that should have already collapsed on
smaller scales at earlier times. Nevertheless this possible numerical
fragmentation is not too much of a concern in the present study,
because even if one was to consider {\em all} the new haloes identified in
the CDM filament at {\em all} redshifts as numerical artifacts, one
still would have to invoke physical fragmentation to explain how the more
massive new WDM haloes formed (see \Fig{totmassfun}).

\subsection{Merger History in Filament} \label{Merger}

\begin{figure}
    \centerline{\psfig{file=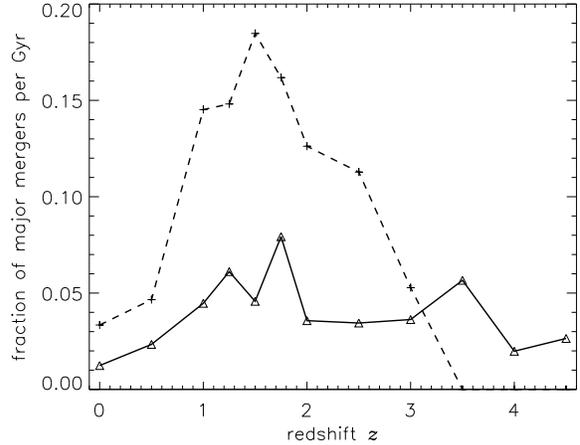,width=\hsize}} 
    \caption{Fraction of merger events per Gyr for which the 
             ratio of the two most massive progenitors is 
             higher than 3:1 (major mergers).}
    \label{merger} 
   \end{figure}

Parallel to fragmentation and creation of new haloes, we also measure
the rate of major mergers within the filaments in both models. A major
merger event is defined by the ratio of the two most massive
progenitors: if this ratio is smaller than 3:1, then the merger is
major, otherwise it is minor.  As before, a lower mass cut-off of
$10^9$\hMsun\ is applied for each halo. \Fig{merger} indicates that at
early times (high redshift) there is no major merger activity in the
WDM filament whereas at later times the relative fraction of major
mergers exceeds that found in the CDM scenario. This supports our
previous claim that haloes at high redshift ($z>3$) do not form
according to the traditional bottom-up hierarchy (i.e. via mergers) in
the WDM filament.  Moreover, since the mass distribution of the
fragments formed through the top-down mechanism between $z=5$ and
$z=2$ is broadly peaked around the Jeans mass (see
\Fig{totmassfun}), one naturally expects the major merger
activity to reach a maximum around $z=2$ in WDM because this epoch
combines a high number density of low mass halos with a small mass
spread in their distribution. In other words, at redshift higher than
2, quite a large fraction of small halos are still forming from
fragmentation, and at lower redshift most of the formation and merging
have already taken place, broadening the mass distribution function.
As a result of these effects, the major merger fraction is strongly
suppressed both at low and high redshifts.  On the other hand, in the
CDM filament, the fraction of major mergers is almost constant in time
and reaches more modest values. This is the result of most of the
haloes being already in place at high redshift ($z>5$) which
guarantees that the mass distribution of haloes will broaden smoothly
over time as merging and accretion proceed.

\section{Discussion and Conclusions} \label{Conclusions}

We presented the first high resolution \nbody\ simulation of the
top-down formation of dark matter halos through the fragmentation of a
dark matter filament. This leads to a number of interesting results:
the epoch of formation of small haloes is pushed back in time to lower
redshifts ($2 < z < 5$) with respect to the standard CDM model, and
the mass distribution of the fragments is broadly peaked around a
Jeans mass of a few $10^9$ M$_\odot$, corresponding to a gravitational
instability of smooth regions with an overdensity contrast around 10
at these redshifts.  We need to undertake detailed hydrodynamic and
star formation simulations to properly assess the impact of such a
mode of formation on dwarf galaxies, but we anticipate the predictions
in terms of clustering and the mean age of the stellar populations to
be quite different from those of the standard CDM model.  We defer
this study to a companion paper.  Reionisation of filamentary
structures might not be a problem since we have shown that massive
objects are already in place at z = 5. Of course, the higher the
redshift the more difficult it will be to reionise the universe unless
the decrease in the number of objects is compensated by an increase in
the efficiency of the ionising sources. Our numerical simulations
strongly suggest that spurious numerical fragmentation may occur in
CDM models due to the softening of the gravity force. However, this is
not to be confused with the real fragmentation observed in the WDM
simulation presented here because in AMR codes such as \mlapm\ the
force resolution is a function of the mass resolution. Therefore, two
haloes with the same number of particles in the \LCDM\ and \LWDM\
simulations are resolved with the same force resolution. This implies
that if we were dominated by artificial fragmentation in both \LCDM\
and \LWDM\, we should not observe different upper mass cut-offs in the
mass distribution functions of new halos. Looking at \Fig{totmassfun}
we clearly see that this is not the case: the mass distribution of new
halos in \LCDM\ drops off sharply at $\approx 2\times 10^9$\hMsun\
whereas it extends to masses $\grtsim 10^{10}$\hMsun\ in
\LWDM. Furthermore, the resolution of our new WDM haloes is high
enough that they should have been detected by the BDM algorithm at
higher redshift had they been present at that time, since CDM haloes
as massive are indeed classified as old !  Finally, we point out that
Melott (1982) already showed that galactic haloes were forming via
real fragmentation in Hot Dark Matter simulations where the resolution
was several orders of magnitudes below what we achieve in this
study. In light of these remarks, we conclude that while our results
for the formation of massive (at least several hundred particles) WDM
haloes through fragmentation are robust, a more detailed study is
needed in order to really quantify numerical fragmentation effects on
smaller scales, as they might be relevant for studies of dwarf
galaxies in cosmological contexts where small haloes are not
sufficiently resolved. More specifically, we believe these numerical
effects will lead to a later formation of low mass haloes in
moderately overdense regions.

\section*{Acknowledgments}
The simulations presented in this paper were carried out on the
Beowulf cluster at the Centre for Astrophysics~\& Supercomputing,
Swinburne University and the Oxford Supercomputer Centre. AK
acknowledges the support of the Swinburne University Research
Development Grants Scheme. JEGD acknowledges enlightening discussions
with James Taylor about granularity, and with Greg Bryan about
numerical resolution issues. The research of JEGD at Oxford is
supported by a major grant from the Leverhulme trust. BKG acknowledges
the support financial support of the Australian Research Council.


\end{document}